\PassOptionsToPackage{svgnames}{xcolor}
\documentclass[preprint,svgnames]{vgtc}               

\graphicspath{{figures/}{pictures/}{images/}{./}} 

\usepackage{times}                     

\usepackage{tabu}                      
\usepackage{booktabs}                  
\usepackage{lipsum}                    
\usepackage{mwe}                       
\usepackage{amsmath}                    
\usepackage{mathptmx}                  
\usepackage{graphicx} 
\usepackage{epstopdf}
\usepackage{svg}
\usepackage{multirow}
\usepackage [english]{babel}
\usepackage [autostyle, english = american]{csquotes}
 \usepackage[final]{changes} 

\newcommand{\ADD}[1]{\added[id={+}]{#1}}
\newcommand{\DEL}[1]{\deleted[id={-}]{#1}}

\MakeOuterQuote{"}

\onlineid{1032}

\vgtccategory{Research}

\vgtcinsertpkg

\ieeedoi{https://doi.org/10.1109/VR59515.2025.00092}

\title{MRUCT: Mixed Reality Assistance for Acupuncture \\Guided by Ultrasonic Computed Tomography}

\author{
Xinkai Wang\thanks{e-mail: xkwang@seu.edu.cn. Joint First Author.}\\ %
        Southeast University %
        \and
Yue Yang\thanks{e-mail: yueyang1@stanford.edu. Joint First Author.}\\ %
     Stanford University %
\and Kehong Zhou\\ %
     Southeast University %
\and Xue Xie\\ %
     Shanghai Sixth People's Hospital %
\and Lifeng Zhu \thanks{e-mail: lfzhulf@gmail.com. Corresponding Author.}\\ %
     Southeast University %
\and Aiguo Song \\ %
     Southeast University %
\and Bruce Daniel \\ %
     Stanford University %
     }

\teaser{
  \centering
  \includegraphics[width=\linewidth]{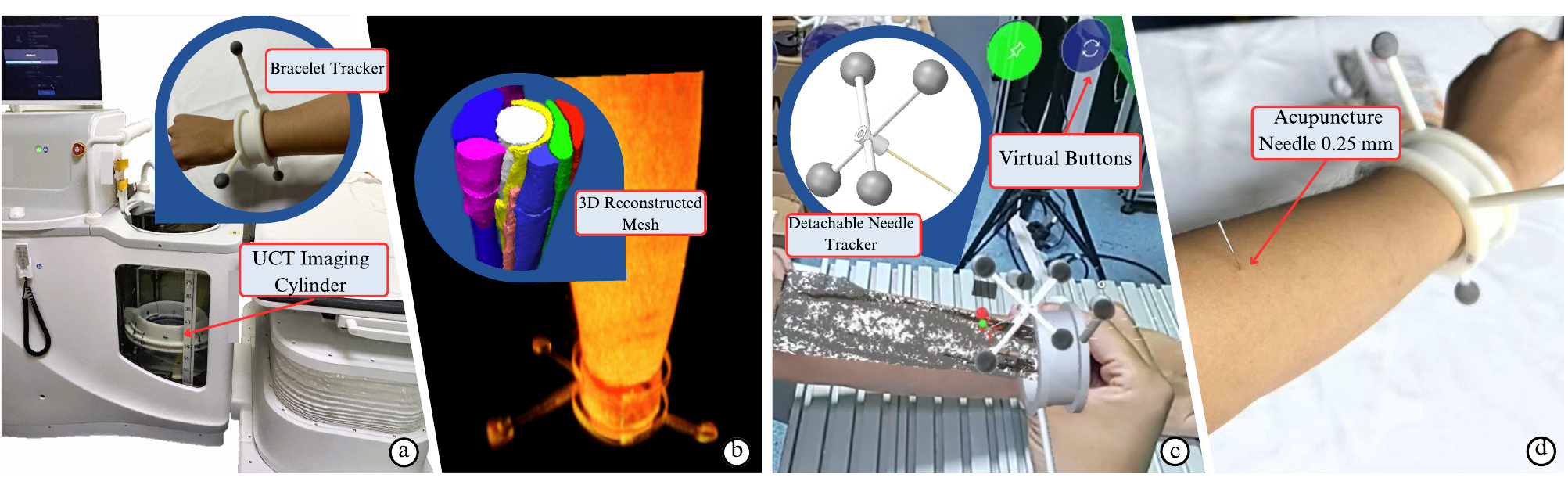}
  \caption{MRUCT is a mixed-reality (MR) acupuncture guidance system that integrates ultrasonic computed tomography (UCT) and an attention-adaptive 3DUI for accurate ultrathin needle insertion. (a) The user places their arm inside a UCT machine while wearing a bracelet tracker to scan the arm. (b) MRUCT reconstructs an accurate 3D representation of anatomical structures from raw, noisy 2D images using an image registration algorithm. (c) Our MR system spatially overlays the virtual model onto the patient's arm and guides needle insertion using needle tracking and an adaptive 3DUI. (d) The needle tracker is detachable and is removed after insertion, ensuring a consistent procedure.}
  \label{fig:teaser}
}

\abstract{
     Chinese acupuncture practitioners primarily depend on muscle memory and tactile feedback to insert needles and accurately target acupuncture points, as the current workflow lacks imaging modalities and visual aids.
     Consequently, new practitioners often learn through trial and error, requiring years of experience to become proficient and earn the trust of patients. Medical students face similar challenges in mastering this skill.
     To address these challenges, we developed an innovative system, MRUCT, that integrates ultrasonic computed tomography (UCT) with mixed reality (MR) technology to visualize acupuncture points in real-time. This system offers offline image registration and real-time guidance during needle insertion, enabling them to accurately position needles based on anatomical structures such as bones, muscles, and auto-generated reference points, with the potential for clinical implementation. 
     In this paper, we outline the non-rigid registration methods used to reconstruct anatomical structures from UCT data, as well as the key design considerations of the MR system. We evaluated two different 3D user interface (3DUI) designs and compared the performance of our system to traditional workflows for both new practitioners and medical students. The results highlight the potential of MR to enhance therapeutic medical practices and demonstrate the effectiveness of the system we developed.
} 


\begin{document}


\firstsection{Introduction}

\maketitle

Acupuncture is a widely practiced therapy in Traditional Chinese Medicine (TCM), known for its ability to alleviate various types of pain, including headaches, arthritis, and muscle pain \cite{kelly2019acupuncture}. Moreover, it serves as an adjunctive treatment, complementing other medical therapies in managing various conditions, such as reducing side effects from chemotherapy or surgery \cite{liu2021neuroanatomical}. Physiological variations between patients make precise preoperative localization of "acu-points" on the human body for needling a significant challenge \cite{sun2023design}. Furthermore, the identification of acupuncture points in traditional procedures often relies on the practitioner’s experience, which can greatly affect the treatment’s efficacy.

To perform acupuncture, practitioners insert thin needles to reach and stimulate "acupoints" which are located near the skin or at varying depths \cite{kim2017acupuncture}. The ability to accurately reach acupoints is critical for effective acupuncture treatment \cite{zhang2021interpretation}. Although acupuncture is generally considered safe, a wide range of adverse outcomes, including trauma and nervous system injury, can occur with improper needle insertion \cite{white2004cumulative}. However, the current clinical workflow lacks imaging modalities to visualize acupoints, largely due to the challenges of referencing imaging results during needle insertion \cite{van2018acupuncture}. As a result, years of training and practice are required to memorize the locations of the 361 standardized acupoints on the human body and to insert needles using experience and the \emph{Finger-Cun} method, which relies on using finger widths to estimate acupoint locations \cite{world2008standard, kim2014positioning}. Despite extensive training, accurately and consistently locating acupoints remains highly challenging, with significant variation in insertion location and depth \cite{lin2013exploration, godson2019accuracy}. Medical students face even greater challenges due to their limited experience with the \emph{Finger-Cun} method\footnote{The \emph{Finger-Cun} method is a traditional technique in TCM used to measure distances on the body, particularly for determining acupuncture points or locating anatomical landmarks.}.

One potential solution is using anatomical landmarks to identify and locate acupoints, which is widely regarded as the most accurate approach \cite{godson2019accuracy}. However, expert interviews revealed that additional training is necessary for acupuncturists to interpret acupoints from 2-dimensional imaging results and effectively apply these findings in practice, complicating the implementation.

To address these challenges, we propose \textbf{MRUCT}, a mixed reality system that utilizes ultrasonic computed tomography (UCT) \cite{gemmeke20073d,lasaygues2011ultrasonic,wei2024clinical} results to provide visual guidance for practitioners to locate acupoints. MRUCT is valuable for acupuncturists in three main aspects: (1) visualizing anatomical structures (bones and muscles) in situ to provide anatomical landmarks; (2) providing software-generated acupoints and insertion trajectories as reference points for manual adjustments; (3) offering real-time guidance through a novel 3D user interface (3DUI) in mixed reality (MR) for needle insertion after trajectory adjustments. We assess MRUCT’s performance through user studies involving inexperienced medical students and experienced acupuncturists, comparing it with a two-ring 3DUI designed by previous research for MR needle insertion guidance and the traditional \emph{Finger-Cun} method \cite{hzhang2024strighttrack, hata2016body, jiang2023wearable}. Accordingly, our work contributes the following:

\begin{itemize} 
    \item A novel MR system that provides real-time guidance for acupuncturists to insert needles accurately, enhancing the effectiveness of acupuncture treatments.
    \item A UCT data processing and offline non-rigid registration workflow that reconstructs accurate 3D representations of patients' anatomical structures from raw 3D ultrasound images. This enables the visualization of bones and muscles, as well as the generation of reference acupoints and trajectories for the MR system to spatially overlay onto the patient.
    \item Design considerations and insights for developing an adaptive 3DUI focused on providing guidance for ultrathin needle insertion.
    \item A comprehensive user study that evaluates the usability of MRUCT compared to the two-ring 3DUI and the traditional \emph{Finger-Cun} method, demonstrating our system’s effectiveness in ease of use, needle insertion accuracy, and consistency.
\end{itemize}

\section{Related work}
\subsection{Imaging Modalities Integrated to Understand and Locate Acupoints}
Researchers, including Leow et al. and Park et al., explored the use of ultrasound to locate various acupoints and demonstrated the potential of 2D ultrasound-guided techniques for acupuncture needle insertion \cite{leow2017exploring, leow2017ultrasonography, park2011acupuncture}. Kim et al. developed a procedure to acquire 2D ultrasound images using a probe to enhance acupuncture safety by visualizing the underlying structures beneath the skin \cite{kim2017development}. However, the lack of spatial information from 2D images and the challenge of managing a probe during needle insertion remains significant, as prior studies primarily focused on visualizing acupoints for anatomical understanding rather than providing real-time guidance. Our system overcomes the challenge of locating acupoints in space from 2D images by adopting UCT and MR technology, enabling the visualization of 3D reconstructed anatomical structures directly on the patient’s body (right arm). 

Magnetic Resonance Imaging (MRI) and Computed Tomography (CT) have also been utilized retrospectively to assess safe needling depths \cite{chou2015retrospective, yang2015safe, chen2009therapeutic}. Lin et al., in their review, noted considerable inconsistency in defining safe needling depths across research groups, with no conclusive evidence linking or quantifying acupoint depth to anatomical structures based on imaging results \cite{lin2013exploration}. Therefore, our system focused on providing insertion trajectory guidance, allowing practitioners to manually adjust the auto-generated insertion trajectory using their hands while wearing an MR head-mounted display (HMD). The insertion depth is not predefined by the system but is visualized through holographic renderings of bone, muscle, and tracked needles.

\subsection{\ADD{Real-Time Anatomical Visualization and 3DUI in AR/MR Systems for Surgery and Needle Guidance}}
\ADD{AR and MR technologies are revolutionizing surgical precision by superimposing real-time anatomical models onto patients, enhancing spatial awareness and reducing reliance on external imaging \cite{moga2021augmented}, \cite{eom2022ar}, \cite{ma2018moving}. These systems enable more precise interventions in various medical fields.}

\ADD{Eom et al. \cite{eom2022ar} developed an AR system for ventriculostomy, improving catheter placement by projecting 3D ventricular models onto the cranial surface. Ma et al. \cite{ma2018moving} introduced an AR navigation system that compensates for motion, optimizing surgical workflows. In orthopedic surgery, Hu et al. \cite{hu2024artificial} used AI-driven AR for knee replacement, enhancing implant alignment with CT scans, while Penza et al. \cite{penza2023augmented} applied AR for robot-assisted surgeries, integrating endoscopic cameras with 3D anatomical models.}

\ADD{In needle guidance, Varble et al. \cite{borde2024smart} demonstrated a goggle-based AR system for needle placement, showing comparable accuracy to traditional ultrasound methods in radiology. Wang et al. \cite{jiang2023wearable} developed a lumbar puncture system combining wearable ultrasound with HoloLens-based AR for improved needle visualization. Kuzhagaliyev et al. \cite{kuzhagaliyev2018augmented} applied AR for pancreatic ablation, improving needle alignment, while Smith et al. \cite{heinrich20222d} found that 3D AR reduced cognitive load and enhanced hand-eye coordination in needle procedures.}

\ADD{Our system focus on real-time acupuncture guidance with UCT modality, visualizing insertion trajectories on the patient’s body and improving precision and situational awareness.}

\subsection{Extended Reality (XR) Systems for Acupuncture Guidance}
From initial virtual representations of acupoints to more recent advancements in Virtual Reality (VR), multiple VR systems have been developed primarily for acupuncture teaching and training \cite{zhang2024acuvr, liang2021analysis, chen2019application}. To enhance the immersion of simulation and training experiences, Sun et al. developed a MR training simulator that allows practitioners to develop muscle memory \cite{sun2023design}. Chen et al. created an AR system on an Android smartphone to visually represent acupoints in space \cite{chen2017localization}, while Zhang et al. introduced an Android app utilizing a new method to localize AR acupoints \cite{zhang2022faceatlasar}. Despite these advancements, significant gaps remain in applying extended reality systems to clinical settings, particularly in working with real human bodies and supporting hands-on, real-world needling practices \cite{zhang2023vr}. To our knowledge, we developed the first HMD-based MR system that provides real-time guidance for acupuncture needle insertion and evaluated its effectiveness on actual patients.

\section{Formative Study}
Previous research has indicated that integrating imaging modalities with XR technologies aids in locating acupoints. However, it remains unclear what design choices should be made and what expectations acupuncture practitioners have for an assistive MR system in clinical environments. This formative study aims to: (1) validate the needs of practitioners for a guided acupuncture procedure; (2) identify challenges they face in daily practice; and (3) derive insights into design choices for a well-integrated, user-friendly, and effective system for clinical applications on real patients.
\subsection{Procedure}
We enrolled 13 participants for the formative study, which included interview sessions and expert discussions. We first conducted semi-structured interviews with seven experienced acupuncturists, asking them to describe the challenges encountered in clinical acupuncture practice (e.g., "How do you decide where to insert needles?" and "Is it challenging to choose the path for needle insertion?") and their expectations for assistive technology providing visual guidance (e.g., "How do you feel about relying on an auto-generated insertion path?"). Researchers conducted face-to-face interviews with each acupuncturist at different acupuncture clinics in two countries. 

\begin{figure}[!t]
    \centering
    \includegraphics[width=\linewidth]{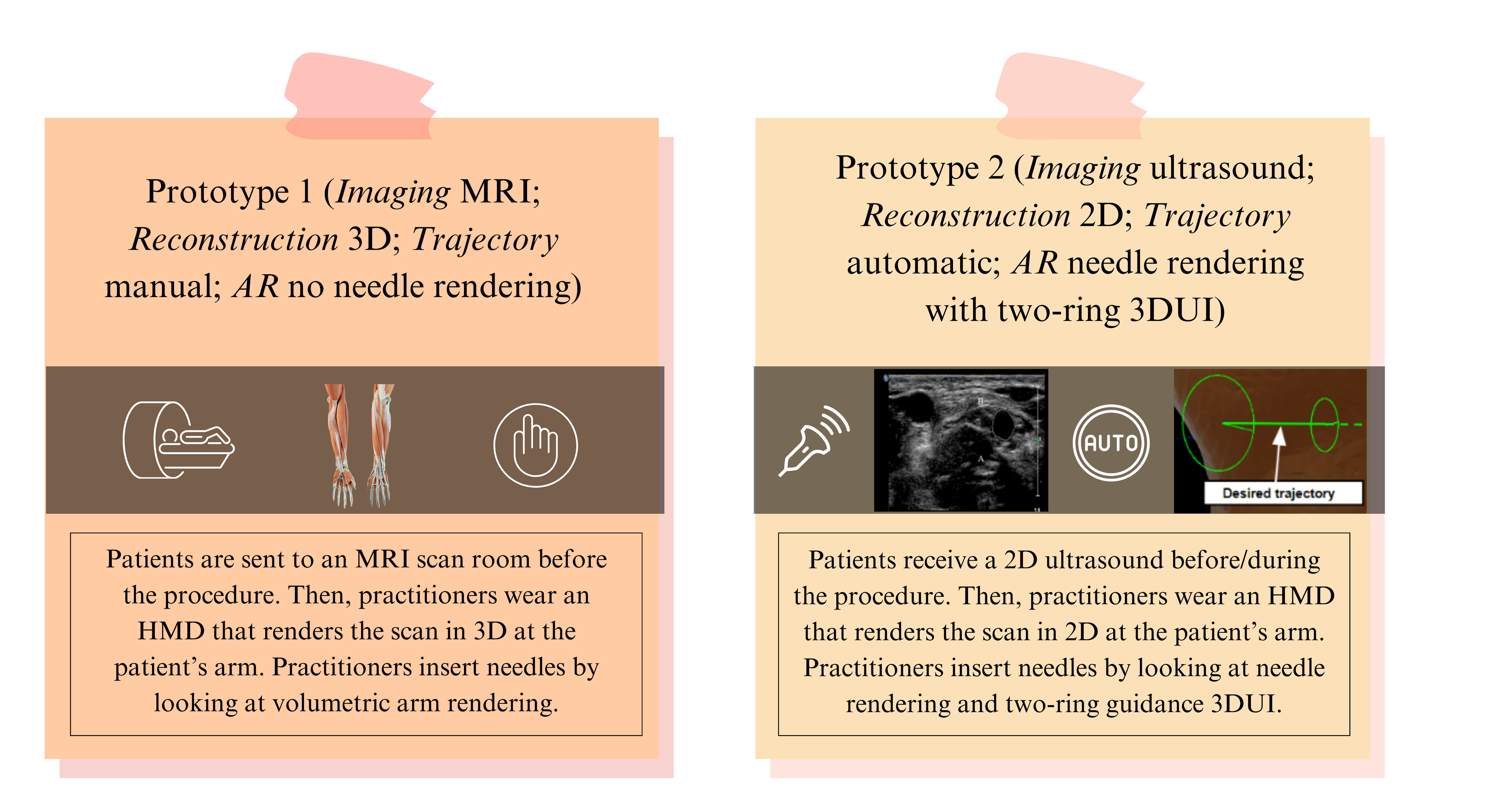}
    \caption{Design cards illustrating different prototype design choices were used, with a short description of the system workflow attached to each card to facilitate understanding and discussion. Note that the two-ring UI and 2D ultrasound images were adopted from previous studies \cite{hzhang2024strighttrack, leow2017exploring}.}
    \label{fig:formative}
    \vspace{-1.5em}
\end{figure}

With the interview data collected, we hosted an expert brainstorming panel consisting of three senior XR researchers (R1-R3), two experienced radiologists (R4, R5), and one acupuncturist (R6). The discussion focused on making design choices for an MR system and integrating imaging modalities. We provided "design cards" in random order, featuring different combinations of imaging modalities, trajectory generation methods, and MR 3DUI, representing prototype designs to address acupuncturists' problems and meet their expectations (see \cref{fig:formative}). Experts were asked to provide feedback on each prototype and explain their preferences for certain methods over others. The feedback and explanations were transcribed and analyzed using the affinity diagram approach \cite{lucero2015using}. Following stages of creating sticky notes, clustering notes on the wall, and walking through them, two researchers grouped feedback based on critiques and preferences to ground discussions and make design choices through an iterative process.

\subsection{Findings}
We present the following qualitative findings from the \DEL{formative} study:
\subsubsection{Imaging Modalities}
Experts discussed three imaging modalities—MRI, CT, and ultrasound—to address the difficulty of locating acupoints. Discussions on 2D and 3D imaging displays were also undertaken to determine the best way to present relevant information in a user-friendly manner. MRI was generally less favored due to its higher cost and longer duration, as expressed by R6, “I don’t think I want to use MRI before acupuncture because it takes too long. And you have to go to a separate scan room and come back.” R4 added, “I don’t think acupuncture needles are MRI-safe because they are metallic. You maybe have to customize MRI-safe needles for it to work.” R1 and R3 noted the preference for 3D rendering and display for its user-intuitive and friendly interface, while R2 raised concerns about the accuracy and validity of imaging registration for 3D display. This suggests that a low-cost and rapid imaging method containing 3D information should be employed before acupuncture.

\subsubsection{Needle Insertion Trajectory and Depth Generation}
In the design cards, we proposed three trajectory generation methods: fully automatic based on anatomical structure analysis, fully manual by practitioners, and a combination of both. Experts preferred the combination approach, which allows acupuncturists to manually adjust trajectories based on auto-generated reference points. This method was favored for its better real-world application, acknowledging the significant variations among patients and the fact that not all acupoints correlate well with anatomical structures.

\subsubsection{Design Insights for MR System}
To make the MR guidance system intuitive and user-friendly, particularly for guiding acupuncture needle insertion, various MR design considerations were discussed. Participants provided insights on how different elements should be rendered in MR. R1 commented, "If you only render the bones and muscles in MR, it would be almost impossible to see your actual needle because occlusion is a major problem, and the needle is very thin." R2 and R3 noted that the thin acupuncture needle poses challenges for tracking and visualizing as a hologram, a widely adopted method to reduce occlusion. R3 added, "I think the acupuncture needle is very unique because it’s super thin and short. Existing 3DUI designs may not work well at this scale." This suggests that both the needle and anatomical structures should be rendered, and the 3DUI must be carefully designed to guide acupuncturists along the AR trajectory.

We summarized the following design considerations that guided the design of our MR system based on findings from our formative study:
\begin{itemize} 
    \setlength{\topmargin}{0pt}
    \setlength{\itemsep}{0em}
    \setlength{\parskip}{0pt}
    \setlength{\parsep}{0pt}
    \item \textbf{DC1}: Use a clinically applicable imaging modality to reconstruct accurate 3D representation of anatomical structures.
    \item \textbf{DC2}: Allow practitioners to adjust auto-generated reference trajectories.
    \item \textbf{DC3}: Render both the acupuncture needle and the patient's anatomical structures in MR.
    \item \textbf{DC4}: Design a 3DUI to clearly and intuitively indicate the distance of the thin needle from the target trajectory.
\end{itemize}

\section{System Design of MRUCT: Enhancing Practitioners' Needle Insertion Accuracy through Volumetric Rendering of Anatomical Structures and Interactive MR Guidance System}
The design of our assistance system, MRUCT, is driven by design considerations from the formative study. This section outlines four key functional components: (1) pre-operative scanning and data processing with UCT-imaging, (2) offline non-rigid registration in the image space, (3) generation of reference trajectory, and (4) real-time MR-based guidance with a novel attention-adaptive 3DUI for ultrathin needle guidance (see \cref{fig:teaser}).
\subsection{Pre-operative Scanning and Data Processing with UCT-imaging}
    MRUCT is personalized for each subject. For new subjects, the appropriate tracking bracelet is selected based on the size of their wrist. The UCT machine then scans the subject's arm to generate a 3D ultrasound image\DEL{ of the arm}, including the tracking bracelet. \DEL{in a process that takes approximately three minutes}\ADD{The scanning takes about three minutes, followed by a two-hour reconstruction phase.} The relative position between the bracelet and anatomical structures \DEL{relevant to acupuncture point localization} (e.g., muscles and bones) is calculated in image space, enabling real-time tracking of virtual anatomical structures in world space (see \cref{fig:teaser}.b).

    To facilitate the subsequent non-rigid registration of UCT images, each acquired 3D ultrasound image (including template data) is preprocessed as follows (see \cref{fig:data processing}):
\begin{figure}[htbp]
\centering
\includegraphics[width=\linewidth]{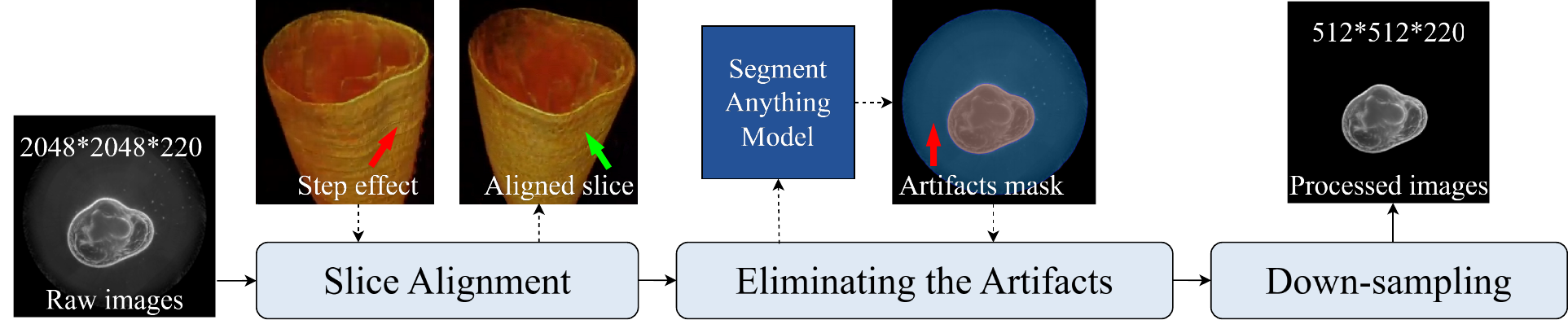}
\caption{Schematic diagram of data processing workflow.} 
\label{fig:data processing}
\vspace{-1.5em}
\end{figure}
\begin{itemize}
    \item \textbf{Slice Alignment:} During UCT scanning, \DEL{subjects may exhibit involuntary movements, resulting in spatial misalignment between neighboring slices and the manifestation of a step effect}\ADD{involuntary movements can cause spatial misalignment between neighboring slices, leading to the step-effect artifact} \cite{lasaygues2011ultrasonic}. To address this, we use the edge feature point-based surface roughness of the 3D ultrasound data as an objective optimization function \cite{liang2018nonrigid}. By adjusting the spatial alignment between slices, we aim to minimize the roughness and improve imaging quality.
    \item \textbf{Eliminating the Artifacts:} \DEL{The heterogeneous acoustic properties of biological tissues}\ADD{Biological tissues} cause ultrasound waves to reflect at tissue interfaces, generating artefacts in UCT images \cite{gemmeke20073d}. To address this, we have adopted the \emph{Segment Anything Model (SAM)} \cite{kirillov2023segment,ma2024segment} to extract the region of interest (ROI) and eliminate background artefacts. \DEL{, thereby enhancing both image quality and the precision of the non-rigid registration algorithm.}\ADD{The pixel center of the UCT image is used as a visual prompt for SAM, with experts manually adding extra points if needed to complete the extraction.}
    \item \textbf{Down Sampling:} The \DEL{UCT }image, originally with a resolution of 2048×2048×220 (0.1mm×0.1mm×1mm), was downsampled to \DEL{a resolution of}512×512×220 (0.4mm×0.4mm×1mm)\DEL{. This procedure accelerates the computational speed of the registration algorithm} \ADD{to speed up the registration process} by lowering computational overhead \cite{amunts2013bigbrain}.
\end{itemize}

\subsection{Offline Non-rigid Registration}
The key to using UCT modalities to assist clinical practitioners in acupuncture lies in segmenting anatomical structures (e.g., bones, muscles) closely related to acupuncture points from the image. This task is challenging and heavily relies on labeled data from medical professionals. In contrast, unsupervised registration networks (e.g., VoxelMorph \cite{balakrishnan2019voxelmorph}, TransMorph \cite{chen2022transmorph}) \DEL{have emerged in recent years but require large-scale paired image datasets, which are time-consuming and labor-intensive to obtain}\ADD{have emerged as promising alternatives. These approaches offer more flexibility by leveraging unpaired image datasets, but they still require large, annotated datasets, which can be difficult to obtain in clinical settings}.

Given the considerable size of the dataset and the need for algorithmic precision, we employed the \emph{Large Deformation Diffeomorphic Metric Mapping (LDDMM)}\footnote{\emph{LDDMM} is a non-parametric registration approach based on principles from fluid mechanics, making it suitable for large deformations.} \cite{beg2005computing} approach for non-rigid registration in image space. These images depict the same anatomy but may exhibit displacement relative to one another.
\ADD{The template image used in the LDDMM registration is a 3D ultrasound image of a specific anatomical structure (such as bone or muscle), which has been labeled by experts while the target image is a preoperative UCT scan of a patient that has not been pre-labeled. The template image serves as a reference atlas representing the anatomy we want to align the patient’s anatomy with.}
Let $I_{m}$ and $I_{f}$ represent the template and target 3D ultrasound images.\DEL{, respectively, from the same area on different patients.} The objective of image registration is to identify the optimal deformation field $\varphi$ that aligns the template image with the corresponding anatomical structure on the target image \cite{sotiras2013deformable}. For a UCT image at a particular time $t$ (where $t \in (0,1)$) during the registration process, we denote it by $I(t)$.

While \emph{LDDMM} does not directly optimize the deformation field $\varphi$, it instead optimizes the spatio-temporal velocity field $v(t,x)$, \ADD{which controls the deformation over time,} and the relationship between them is as follows:
\vspace*{-0.5\baselineskip}
\begin{equation}
    \partial t \varphi (x,t) = v (t,\varphi (x,t))
    \vspace*{-0.5\baselineskip}
\end{equation}

Based on this, we define the following optimization objective:
\vspace*{-0.5\baselineskip}
\begin{equation}
\begin{aligned}
    v^{*} =  & \mathop{\arg\min}\limits_{v}\int_{0}^{1} \left \| v(t) \right \|_{L}^{2}dt + MSE(I(1),I_f)
    \\
    & s.t, \quad I(0)=I_m; \quad \partial t I+\left \langle \nabla I, v \right \rangle = 0 
\end{aligned}
\vspace*{-0.5\baselineskip}
\end{equation}
Here, $\left | v(t) \right |_{L}^{2} = \left \langle Lv, v \right \rangle$, where $L$ is a differential operator used to smooth the velocity field and eliminate fold-over effects in the deformation field \cite{shen2019region}. Additionally, $MSE(A,B)$ is commonly used to assess the similarity between image $A$ and image $B$.

After a \DEL{short} time interval $\delta t$, we obtain the image at the next moment:
\vspace*{-0.5\baselineskip}
\begin{equation}
    I(t+\delta t)=I(t)-\left \langle \nabla I,v  \right \rangle \delta t
    \vspace*{-0.5\baselineskip}
\end{equation}

\ADD{Without non-rigid registration, misalignment between the 3D model and patient anatomy could lead to incorrect needle placement. Template-based registration is essential for accurate needle insertion guidance.}
\subsection{Generation of Reference Trajectory}
Although MRUCT can register specific anatomical structures in image space, it remained challenging for most untrained medical students to directly label acupoints based on muscle and bone locations. To address this, template images were scanned with real needle positions placed by professional acupuncturists in the world space (which were removed as background by SAM during data preprocessing). The entry and exit points of the needle on the UCT image served as labeled needle trajectories. The deformation field acquired during the registration process was then used to transfer the needle trajectory to the target image space, generating reference needle points for practitioners. This allows practitioners to further adjust acupoint locations in accordance with the anatomical structures.

\subsection{Real-time MR-based Guidance}
The implementation of our MR system includes three core components: a bracelet tracker with four retro-reflective markers facing in four directions (see \cref{fig:teaser}.a), a detachable needle tracker (see \cref{fig:teaser}.c), and an optic see-through head-mounted display (Microsoft HoloLens 2). To track the marker arrays, we used the infrared camera on the HoloLens 2 set to short-throw mode, utilizing algorithms from \cite{martin2023sttar} and enhancing tracking stability by incorporating additional frame-to-frame interpolation. A Kalman filter was integrated to minimize noise and reduce jitter in tracking performance, which is particularly beneficial for the bracelet tracker where markers are more widely spaced. This results in a smoother and improved user experience compared to a direct implementation of the tracking algorithm. The Unity 3D Engine and Mixed Reality Toolkit 2 packages are employed to facilitate basic interaction functionalities \cite{polar-kev2024May}. Users can interact directly with our MR system using their hand to adjust reference trajectories as needed and can move to the next trajectory by pressing a virtual button (see \cref{fig:teaser}.c).

To accurately track the positions of the arm and acupuncture needle within the HMD-world coordinates, we developed two trackers with retro-reflective markers. These trackers were manufactured using a CreatBot D600 Pro 3D printer with ABS material, chosen for its lighter weight and greater durability compared to PLA. The arm tracker comprises two parts printed separately and connected with snap-fit joints, allowing for easy assembly and adjustment. We produced four wrist sizes of the arm tracker (150, 170, 180, and 200 mm) to ensure a proper fit for different users. It is important to note that the placement of the retro-reflective markers remains consistent across all sizes to ensure accurate registration of anatomical structures. For the needle tracker, pivot calibration was performed using the Polaris Vega Optical Navigation Solution from NDI, achieving a root mean square error of 0.17 mm.

\begin{figure}[!t]
    \centering
    \includegraphics[width=\linewidth]{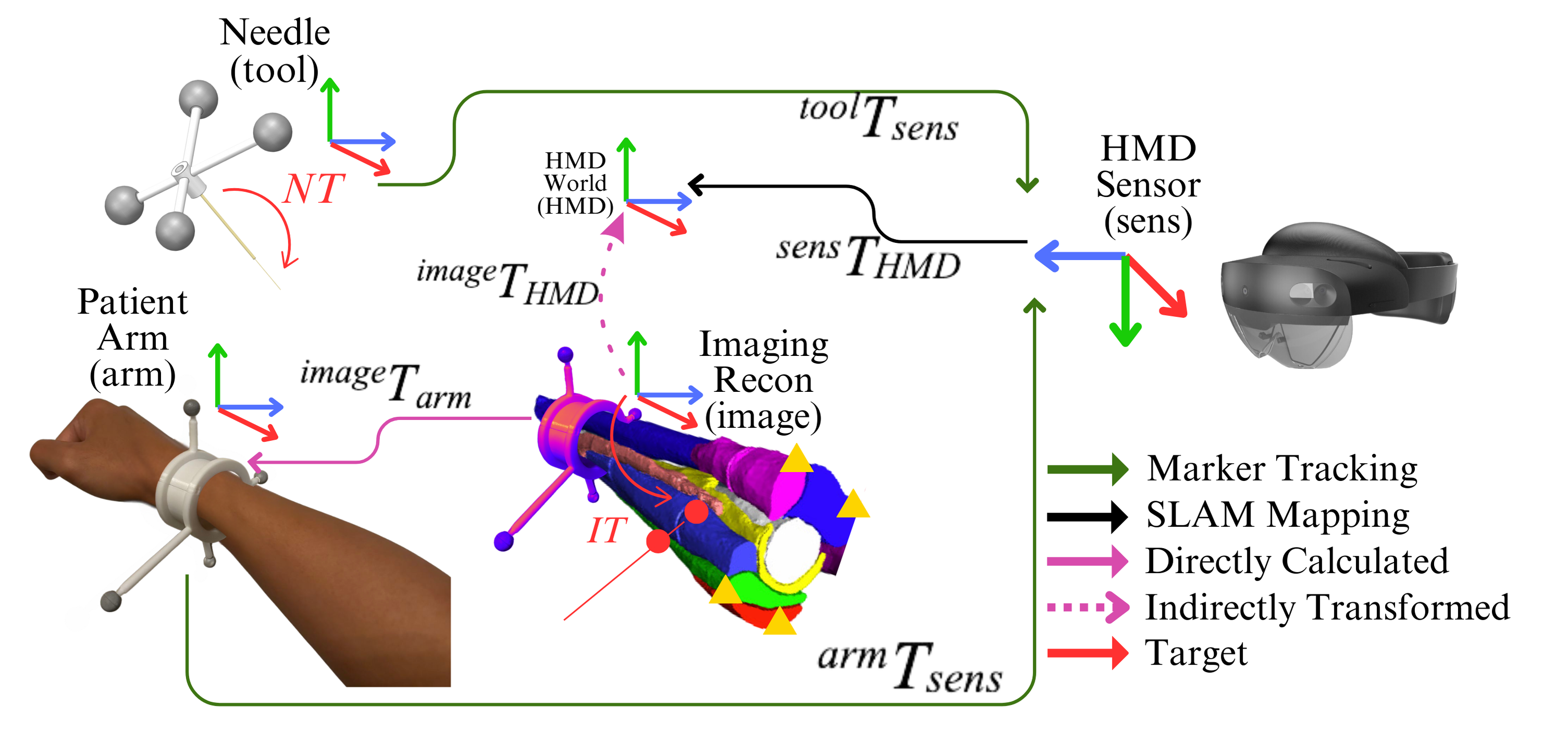}
    \caption{The transformation system of MRUCT. The HMD sensor tracks the patient's arm and needle and provides real-time transformation. The 3D imaging reconstruction is aligned to the patient's arm through local coordinates of retro-reflective mark coordinates and four labeled mesh coordinates (yellow triangles). The reference insertion trajectory is automatically computed from entry and end points (red circles). }
    \label{fig:arregister}
    \vspace{-1.0em}
\end{figure}

The registration workflow facilitates accurate spatial alignment between the UCT imaging space (represented by a 3D reconstructed mesh) and the HMD world space. \ADD{This section explains the coordinate transformations and the underlying computations that map the 3D reconstruction from UCT imaging onto the human arm. The 3D visualization moves and rotates with the bracelet and can be occluded by the tracked and virtually rendered acupuncture needle. Based on findings from our formative study, the real-time tissue deformation on the human arm during the acupuncture procedure is not a significant source of error, primarily due to the ultrathin nature of the acupuncture needle. Therefore, our system rigidly overlays the 3D reconstruction in situ.} In our system (see \cref{fig:arregister}), practitioners aim to align the tool with the pre-defined reference trajectory shown in the imaging. This trajectory is indicated by a red line connecting two red dots, representing the entry and exit points. The goal is to ensure that the needle trajectory $\mathbf{NT}$ and the image trajectory $\mathbf{IT}$ are precisely aligned and registered to the HMD world space. Considering the two red points on image as $\mathbf{p}_{{entry}}$ and $\mathbf{p}_{{end}}$, we have
\vspace*{-0.5\baselineskip}
\begin{equation}\label{equation4}
\begin{aligned}
\mathbf{IT} &= \mathbf{p}_{{entry}} - \mathbf{p}_{{end}} \\
\mathbf{IT} \cdot {}^{\text{image}}T^{}_{\text{HMD}} &= 
\mathbf{NT} \cdot {}^{\text{tool}}T^{}_{\text{sens}} \cdot 
{}^{\text{sens}}T^{}_{\text{HMD}}
\end{aligned}
\vspace*{-0.5\baselineskip}
\end{equation}
\noindent

As in \cref{fig:arregister}, ${}^{\text{tool}}T^{}_{\text{sens}}$ is the transformation from needle tool to HMD sensor space, and ${}^{\text{sens}}T^{}_{\text{HMD}}$ represents the transformation from HMD sensor to HMD world space. ${}^{\text{tool}}T^{}_{\text{sens}}$ is obtained by the needle trackers through the tracking algorithm using the HMD depth sensor. ${}^{\text{tool}}T^{}_{\text{sens}}$ is provided by the native SLAM algorithm on HMD that maps its coordinate to the world coordinate. These two transformations are known. However, ${}^{\text{image}}T^{}_{\text{HMD}}$ represents the transformation from imaging reconstruction to HMD world space, which is unknown because of the additional imaging modality, UCT, we introduced and integrated. Since this transformation is required to register the $\mathbf{IT}$ to HMD world space, we indirectly calculate this transformation by 
\vspace*{-0.5\baselineskip}
\begin{equation}\label{equation5}
\begin{aligned}
{}^{\text{image}}T^{}_{\text{HMD}} &= 
{}^{\text{image}}T^{}_{\text{arm}} \cdot {}^{\text{arm}}T^{}_{\text{sens}} \cdot 
{}^{\text{sens}}T^{}_{\text{HMD}}
\end{aligned}
\vspace*{-0.5\baselineskip}
\end{equation}
\noindent

\begin{figure}[!t]
    \centering
    \includegraphics[width=\linewidth]{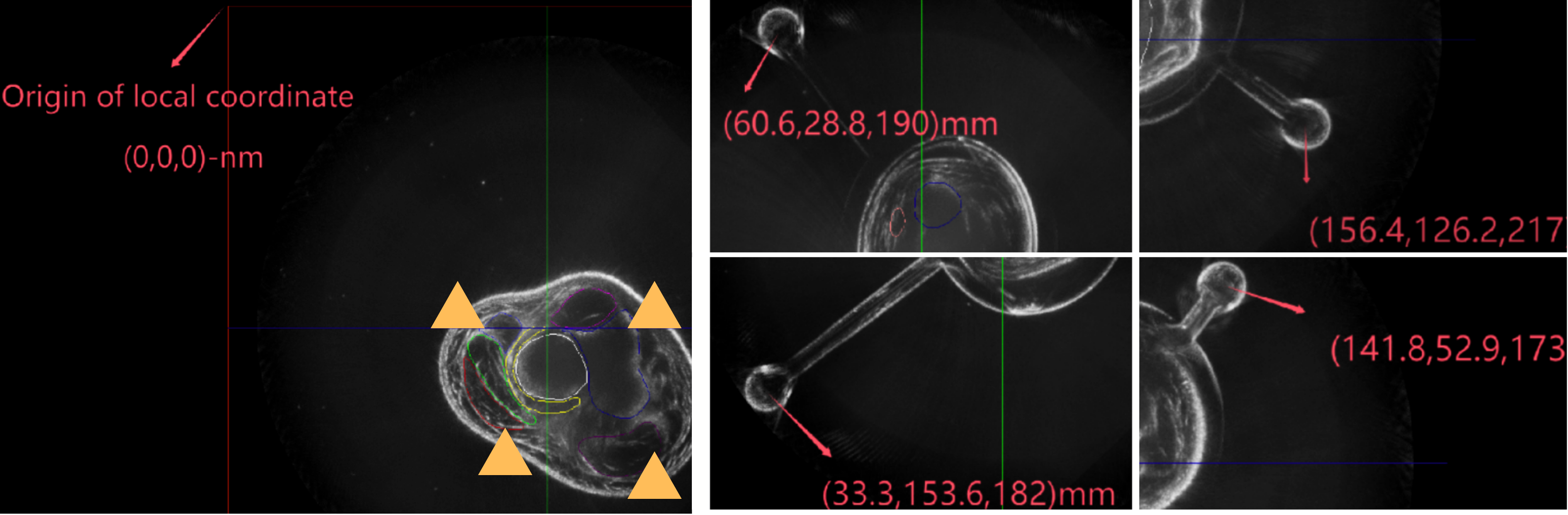}
    \caption{The identification of local coordinates of marker centers and four mesh in imaging space. The coordinates were used to spatially align the 3D reconstructed mesh to the patient's arm position.}
    \label{fig:localcoordinate}
    \vspace{-1.5em}
\end{figure}

The transformation \({}^{\text{arm}}T^{}_{\text{sens}}\) is continuously updated by the tracking algorithm and represents the transformation from the patient's arm to the HMD sensor space. The transformation \({}^{\text{image}}T^{}_{\text{arm}}\) denotes the conversion from the imaging reconstruction to the patient's arm. This is calculated directly using the local coordinates of the center of the retro-reflective markers on the bracelet (see \cref{fig:localcoordinate} right) and the four automatically labeled meshes at the corners of the end image within the imaging coordinates (see \cref{fig:localcoordinate} left and \cref{fig:arregister}). Specifically, we calculate the transformation from the imaging reconstruction to the patient's arm by measuring the relative distances between the four labeled meshes and the center of the markers. Since the bracelet, fitted specifically for each patient, is worn from pre-operative imaging through the intra-operative procedure, this transformation is considered highly rigid and consistent, enabling accurate direct calculations. The accuracy of this transformation is evaluated in \cref{sec:before-human-loop-error}. Thus, our system provides real-time tool tracking and navigation through 
\vspace*{-0.5\baselineskip}
\begin{equation}\label{equation6}
\begin{aligned}
\mathbf{IT} \cdot {}^{\text{image}}T^{}_{\text{arm}} \cdot {}^{\text{arm}}T^{}_{\text{sens}} \cdot 
{}^{\text{sens}}T^{}_{\text{HMD}} &= 
\mathbf{NT} \cdot {}^{\text{tool}}T^{}_{\text{sens}} \cdot 
{}^{\text{sens}}T^{}_{\text{HMD}}
\end{aligned}
\end{equation}
\noindent

\begin{figure}[!t]
    \centering
    \includegraphics[width=\linewidth]{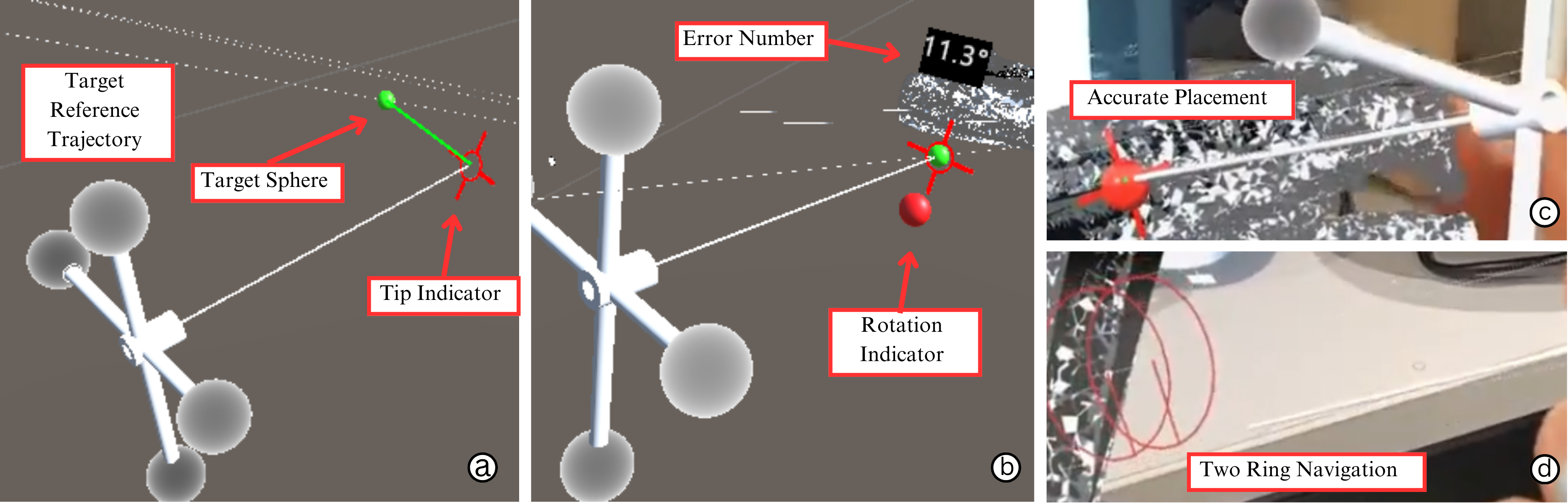}
    \caption{Illustration of MRUCT attention-adaptive 3DUI (a,b,c) and the two ring system (d). (a) shows the mechanism of the tip indicator navigation system. (b) illustrates the activation of the rotation error indicator. (c) demonstrates a precisely placed needle by using MRUCT. (d) is the comparative method}
    \label{fig:uidesign}
    \vspace{-1.5em}
\end{figure}

Guided by DC4, we developed a new 3D User Interface (3DUI) navigation system that assists users in aligning the acupuncture needle with reference trajectories. This system includes a needle tip error indicator and a rotation error indicator. The tip indicator (see \cref{fig:uidesign}.a), designed as a red cross with a central circle, directs users to align their needle with the reference insertion trajectory, aiding in the precise positioning of the needle tip along this line. This indicator is anchored at the needle tip and updates in real-time as the needle is tracked. A vertical green projection line extends from the tip indicator to the reference trajectory, constantly updating to show the optimal path. The intersection of this line with the reference trajectory is marked by a green target sphere, indicating where the needle tip should aim. Users can follow this green line to approach the needle tip to the target sphere. 

The rotation error indicator activates when the needle tip is within 1 cm of the target sphere (see \cref{fig:uidesign}.b). It is rendered as a red sphere and calculated by comparing the HMD world orientation of the needle trajectory with the reference insertion trajectory, signifying the rotational error. This error, quantified in degrees, is displayed above the green target circle and continuously adjusts to face the HMD. Users can rotate the needle to align the red sphere with the green target sphere. 

\ADD{We selected spheres due to their symmetry and lack of orientation dependence. Unlike cones or other primitives that can be visually ambiguous when viewed from certain angles, spheres maintain uniform visibility from any perspective. This attribute aligns with principles discussed by \cite{laviola20173d}, which highlights the importance of intuitive and perspective-independent designs for 3D interfaces. Additionally, spheres can potentially reduce cognitive load for users by eliminating the need to interpret directional cues that cones or arrows might require, which is critical for accurate thin-needle alignment. Red and green colors were chosen due to their strong contrast and ease of perception in controlled environments, as noted by \cite{ware2019information}.}

Our system also provides guidance on insertion depth by rendering a precise representation of the patient’s anatomical structure and continuously tracking and visualizing the needle tip. Users determine the depth by comparing the tip indicator’s position with nearby anatomical landmarks, such as muscles and bones. The system does not include a 3DUI for precise depth navigation because acupoints are considered energetic zones rather than specific points, typically spanning 2-3 mm in diameter. Offering precise depth guidance could potentially confuse users.

Our new guidance has the following novelty compared to previous systems: (1) \textit{Attention-Adaptive Interface.} To reduce mental load and enhance user experience, we adapt the 3DUI visualization based on the user’s focus. Initially, when the user wears the HMD and holds the needle, only the tip indicator navigation is displayed, directing their attention solely to aligning the needle tip with the insertion trajectory. As the needle moves along the insertion trajectory, the 3DUI dynamically updates and shifts its display location to the needle tip, the location of the user's attention when inserting needles (see \cref{fig:uidesign}.c); (2) \textit{Ultrathin Needle Optimizations.} We enhance the visibility of the ultra-thin acupuncture needle by placing a clear red marker at its tip in the HMD world coordinate. This feature addresses the challenge of the needle's thinness and visibility, preventing users from losing track of the needle tip and facilitating accurate insertion.

In our workflow, users initially use the HMD as an assistive tool to sterilize the insertion point on the patient's arm. After the pre-operative procedure, three auto-generated trajectories are displayed in HMD, allowing users to adjust their positions manually if necessary. The needle tip and rotation error indicators then assist in precisely positioning and orienting the needle tip and trajectory to match the reference insertion trajectory. Once the needle contacts the patient's arm, users start to insert the needles and stop based on references to other anatomical structures. We compare our 3DUI with a two-ring system adopted from previous studies \cite{jiang2023wearable, hzhang2024strighttrack}, where the radius of the ring indicates how far the needle trajectory is from the entry or endpoint along the trajectory (see \cref{fig:uidesign}.d). The two rings are fixated at these two points all the time, and users need to move the needle to make both rings smaller. The red line within each ring indicates the direction of movements. 

\section{Evaluation Study}
A series of experiments\footnote{All experiments were carried out in a controlled laboratory environment under clinician supervision \ADD{at the Shanghai Sixth People's Hospital (Shanghai, China)}.} was conducted to investigate three key questions with the objective of validating the effectiveness of the MRUCT:
\begin{itemize}
    \item \textbf{RQ1} —— Can MRUCT, a UCT imaging-based MR system, achieve useful registration of specific anatomical structures (e.g., muscles, bones)?
    \item \textbf{RQ2} —— Can MRUCT help experienced practitioners perform acupuncture more efficiently by improving point location accuracy and speeding up the needling process?
    \item \textbf{RQ3} —— Can MRUCT help medical students with minimal training perform acupuncture tasks correctly?
    \item \textbf{RQ4} —— Can MRUCT outperform other navigation 3DUI and be clinically applicable and easy to use?
\end{itemize}

\subsection{Evaluation of Image Registration}
\label{sec:before-human-loop-error}
To address RQ1, we collected a series of UCT images of the left and right arms from 16 patients (9 males, 7 females) and conducted a comprehensive evaluation of the \emph{LDDMM} model on 32 images pairs.
\begin{figure*}[!t]
\centering
\includegraphics[width=\linewidth]{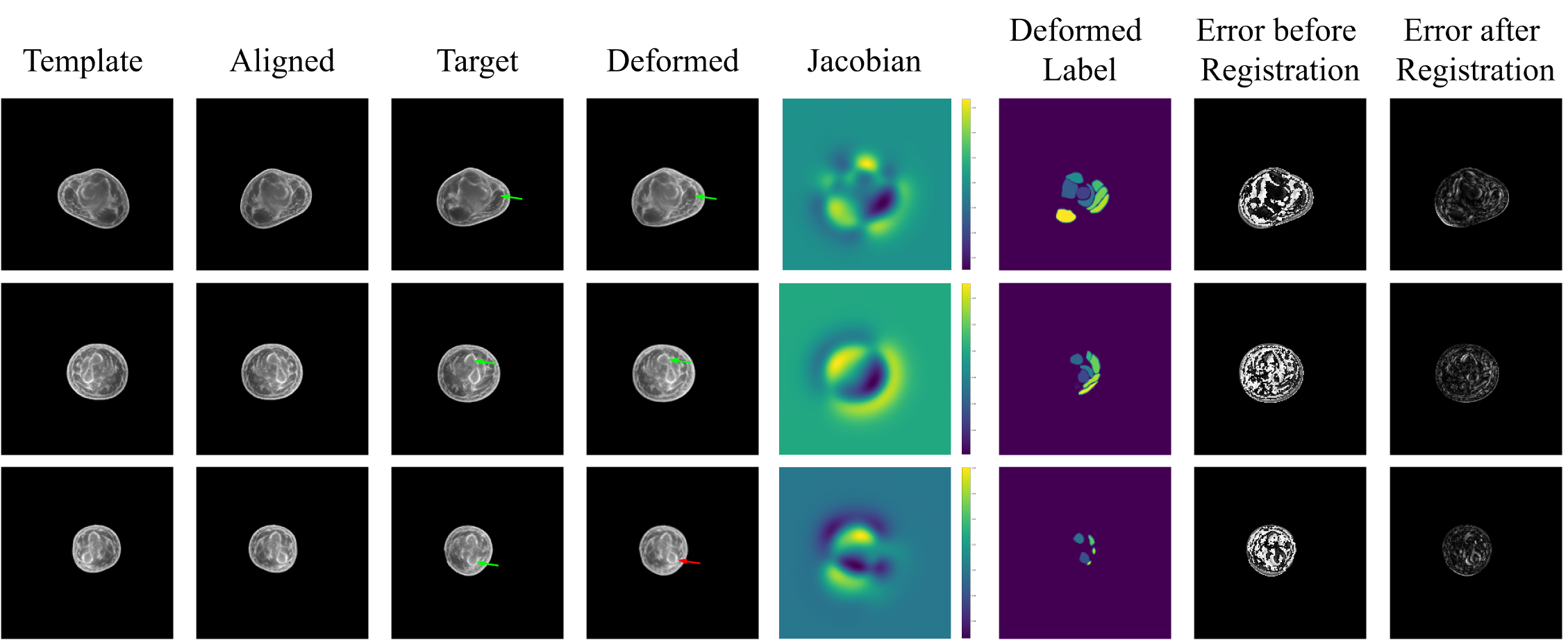}
\caption{\emph{LDDMM} image registration results: Arm images with labeled data (template) are registered to the scanning images of another patient (target). From the first to the third row, we have selected slices 10, 90, and 150 as experimental results for display. Deformations are well-captured in the anatomical region of interest, and the background, along with other regions, is regularized as expected. Columns 1 to 6 show results in both image space and label space, while columns 7 and 8 display the discrepancy between the aligned and deformed images. The 5th column also shows the determinant of the Jacobian of the spatial deformation, $\varphi$.} 
\label{fig:evaluation}
\vspace{-1.5em}
\end{figure*}
\cref{fig:evaluation} shows the results from one of the sample registration experiment pairs, including the template image, the aligned image, the target template, the deformed image, and the discrepancy between the aligned and deformed images. As shown, the discrepancy before and after registration has significantly decreased, while the anatomical structures of the template image remain largely consistent. We evaluated the Dice score between the deformed bone and the target bone (radius and ulna), achieving an average of $89.71\%$ across all pairs. Similarly, we obtained a Dice score of $72.33\%$ between the deformed muscle and the target muscle (e.g., anconeus and supinator). The difference between the scores is due to the fact that bone is clearly imaged in the UCT modality, whereas muscle has a significantly less well-defined border compared to bone \cite{wei2024clinical}.

\cref{fig:evaluation} also shows the determinant of the Jacobian of the deformation map, $J_{\varphi}(x) = \left | D\varphi \right |$, defined in the template image space. The area of the anatomical structure of interest exhibits large deformations, while other regions are volume-preserved (with values close to 1). Overall, most deformations are smooth.

\subsection{Evaluation of End-to-End System Error}
Before evaluating the system performance on real patients used by actual medical students and acupuncturists, we evaluated the end-to-end tracking accuracy of MRUCT. At this stage, the accuracy of MRUCT, excluding any human error, depends on four error sources: the HMD SLAM error, marker tracking error, pivot calibration error, and UCT registration error. 

We initially inserted six acupuncture needles into the patient's arm and captured UCT images. The entry points of these needles were manually marked as landmarks for subsequent error analysis. Using our pivot-calibrated needle, we precisely positioned the real-world needle tip at the six designated entry points, denoting the needle tip coordinates as $\mathbf{TC}$ and the landmark coordinates in the imaging space as $\mathbf{LC}$). Employing the workflow and annotations outlined in (see \cref{fig:arregister}), we derive the following equation:
\vspace*{-0.5\baselineskip}
\begin{equation}\label{equation6}
\begin{aligned}
\mathbf{LC} \cdot {}^{\text{image}}T^{}_{\text{arm}} \cdot {}^{\text{arm}}T^{}_{\text{sens}} \cdot 
{}^{\text{sens}}T^{}_{\text{HMD}} &= 
\mathbf{TC} \cdot {}^{\text{tool}}T^{}_{\text{sens}} \cdot 
{}^{\text{sens}}T^{}_{\text{HMD}}
\end{aligned}
\vspace*{-0.5\baselineskip}
\end{equation}
\noindent

and the error term $\mathbf{\epsilon_{system}}$ as:
\begin{normalsize}
    \begin{equation}\label{11}
\mathbf{\epsilon _{system}} = \left |\mathbf{LC} \cdot {}^{\text{image}}T^{}_{\text{arm}} \cdot {}^{\text{arm}}T^{}_{\text{sens}} \cdot 
{}^{\text{sens}}T^{}_{\text{HMD}} - \mathbf{TC} \cdot {}^{\text{tool}}T^{}_{\text{sens}} \cdot 
{}^{\text{sens}}T^{}_{\text{HMD}} \right |
\end{equation}
\end{normalsize}

We calculated the error result for each skin landmark and repeated the same process again for another patient. The average error from 12 data points was 0.32 mm $\pm$ 0.09 mm.

\subsection{Evaluation of System Usability and Overall Performance}
\subsubsection{Experimental Procedure}
We recruited 31 participants, including two expert-level acupuncturists with over 10 years of experience, one novice acupuncture practitioner, and 28 medical students. After reviewing and signing a consent form and completing a demographic survey, all participants took part in the study. \ADD{The study was performed under the oversight of the Ethics Committee of Shanghai City Sixth People's Hospital (Shanghai, China) and the consent from the human subjects in the research was obtained.} Only one patient (P1) was enrolled, and participants inserted needles at three targeted acupoints on P1’s right arm (LI 10 Shousanli, LI 8 Xialian, and LI 7 Wenliu). Prior to the study, P1 underwent pre-operative UCT imaging while wearing the best-fitting bracelet. \DEL{The imaging procedure, which took approximately three minutes, was performed inside a UCT machine, and the raw images}\ADD{The raw UCT images} were processed on a PC equipped with 32GB RAM, an NVIDIA GeForce RTX 3090 with 24 GB, and an Intel i9\ADD{-10900k 5.30GHz} processor. A 3D reconstruction of P1’s arm and reference insertion trajectories were generated, and spatial registration was completed as described in the registration workflow.

The first participant, an experienced acupuncturist, validated and made minor adjustments to the auto-generated trajectories using the HMD. These adjusted trajectories were then used for all subsequent participants. Each participant underwent a training session to familiarize themselves with both MRUCT and the two-ring navigation system, and the study procedures were explained in detail. Participants performed two insertions per trajectory using both MRUCT and the two-ring system, with the sequence randomized. The needle remained inserted in P1’s arm while the needle tracker was moved to the next needle. Minimal interference between the two insertions was assumed due to the ultrathin nature of the acupuncture needles. This process was repeated for all three trajectories, resulting in a total of six needles being inserted into P1’s arm. The two experienced acupuncturists also performed needle insertions using the traditional figure-cun method without MR technology to facilitate a comparison with MR-assisted insertions for answering RQ2.

After the insertions, participants completed the System Usability Scale (SUS) and NASA Task Load Index (TLX) questionnaires, along with an additional questionnaire assessing overall system performance, clinical applicability, learning barriers, and providing space for additional feedback. They were also asked to rank their preferred 3DUI system and explain their choice, highlighting specific advantages. A post-UCT scan of P1’s arm was conducted for six medical students and two acupuncturists to facilitate subsequent quantitative analysis for RQ2 and RQ3.

\subsubsection{Evaluation Metrics}
To answer RQ4, system effectiveness was qualitatively evaluated using questionnaire results. Non-parametric statistical comparisons were made using the Wilcoxon signed-rank test, as the Shapiro-Wilk test indicated non-normal distribution of data.

Quantitative evaluation of overall system performance involved reconstructing imaging results from the post-UCT scan. The patient’s arm with six inserted needles was reconstructed, and actual needle entry and endpoints were recorded. The three reference trajectories in HMD, manually adjusted by the first acupuncturist, served as the ground truth. Ground truth entry points were directly derived from these reference trajectories, while endpoints were manually labeled by a second experienced acupuncturist. Additionally, a circular region with a 1.5 mm radius around each HMD reference point was delineated to represent the optimal acupoint area. Consequently, the insertion error is defined as the deviation—the shortest distance between the inserted point and the ground truth point—minus the radius of the optimal region.

To answer RQ2, reconstructed entry and endpoint data using the figure-cun method were compared to the HMD reference points to calculate the figure-cun insertion error. For each entry point comparison, the straight-line distance between the two points was calculated, and for the endpoint, the distance between the edge of the circle around the HMD endpoint and the reconstructed endpoint was calculated and considered the endpoint error. This allowed a comparison of insertion-using-HMD errors to insertion-using-figure-cun errors, helping us to evaluate how MRUCT assists experienced acupuncturists in achieving greater accuracy. 

For RQ3, insertion-using-HMD errors were calculated for medical student participants, which helped evaluate system performance on individuals with minimal clinical training.

\subsubsection{Qualitative Results}
Study participants (17 males and 14 females) were recruited from two universities across two continents. Their ages ranged from 19 to 59 years old, with a mean age of 29.9 and a standard deviation of 11.9. Participants reported moderate experience with mixed reality systems, averaging 2.3 (SD = 0.7) on a Likert scale from 1 (minimal) to 4 (high). Similarly, experience with acupuncture averaged 2.7 (SD = 0.8).

\begin{figure*}[h]
    \centering
    \includegraphics[width=0.7\linewidth,height=1.8in]{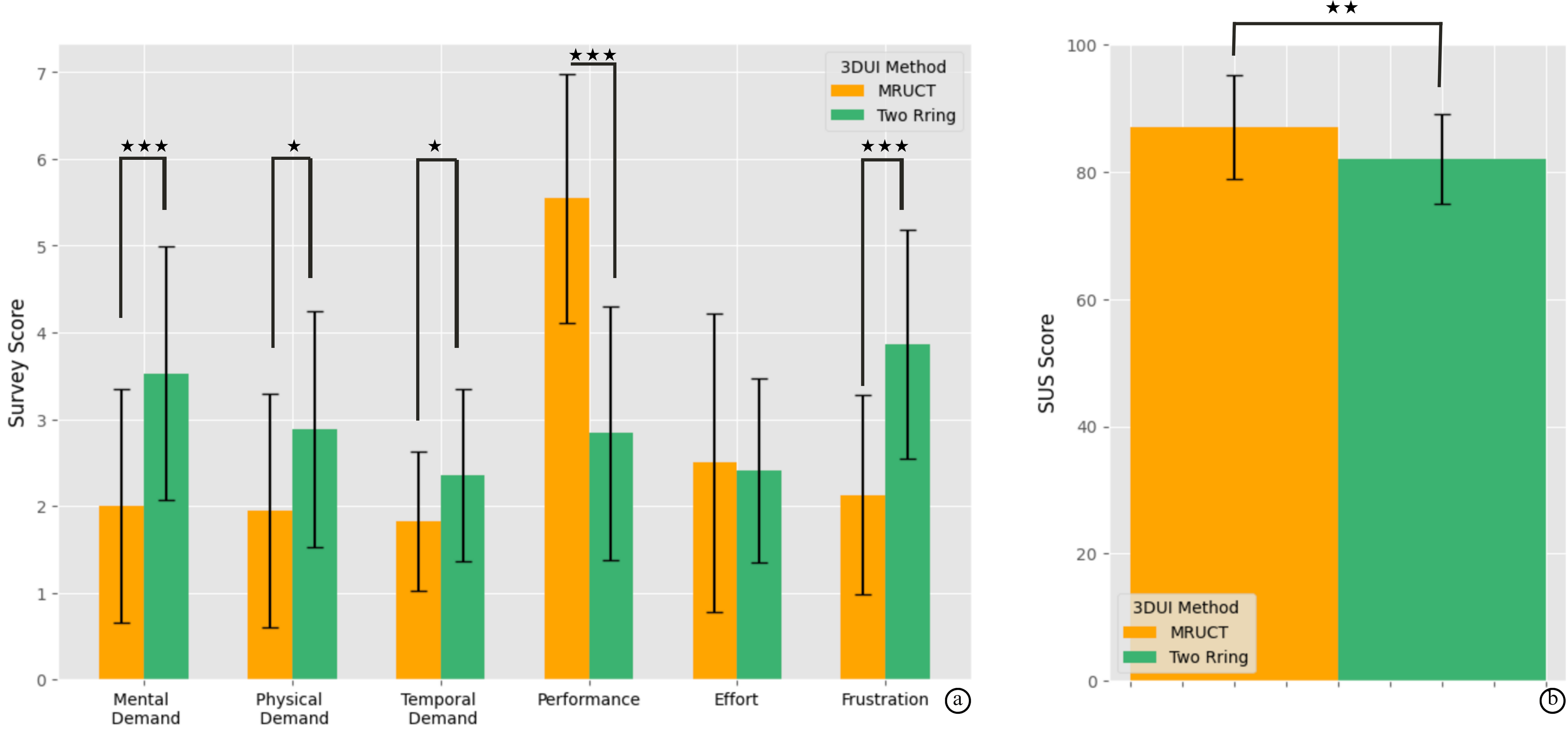}
    \caption{Qualitative survey results from NASA-TLX and SUS questionnaires. (a) shows the comparative analysis of two 3DUI methods in six measuring scopes. (b) is the direct comparison of SUS scores. }
    \label{fig:eval_qual}
    \vspace{-1.5em}
\end{figure*}

Data collected from all 31 participants were analyzed comparatively (see \cref{fig:eval_qual}.a). Utilizing a 7-point Likert scale, the NASA Task Load Index (TLX) indicated that the MRUCT system resulted in a statistically lower mental demand compared to the two-ring method (Z = 3.4, p  \textless .001), suggesting that MRUCT is easier to use. Participants also experienced significantly less frustration with MRUCT compared to the two-ring system (Z = 3.7, p  \textless .001). This reduction in frustration might be attributed to the attention-adaptive design of our system, as participants reported challenges in simultaneously managing and focusing on the two rings. Furthermore, the self-rated performance score was significantly higher for MRUCT (Z = -4.3, p  \textless .001), reflecting a notable improvement in self-confidence and performance evaluation. The following quantitative results from actual insertion evaluation validate this finding.

Additionally, the physical and temporal demands associated with using MRUCT were significantly lower (Z = 2.5, p = .012; Z = 2.3, p = .019), indicating that participants generally felt less rushed and less fatigued when using our system. No statistically significant differences were observed in effort levels, suggesting that comparable efforts were exerted to achieve performance outcomes.

Using a 1-5 Likert scale, the SUS scores were calculated for each participant and 3DUI system (see \cref{fig:eval_qual}.b), utilizing a customized version of the standard SUS questionnaire tailored to our study context. Questions were specifically adjusted to better reflect our scenario (e.g., “I would imagine that most acupuncture practitioners would learn to use this system very quickly.”). The MRUCT \DEL{system}achieved significantly higher SUS scores than the two-ring \DEL{system}(Z = -2.6, p = .008), demonstrating its high usability in clinical settings.

We asked participants their preference of using MRUCT or the two-ring navigation system, and an overwhelming majority—30 out of 31—opted for MRUCT. The participant who chose the two-ring system cited his familiarity with a similar MR system previously used for guiding K-wire insertion, which influenced his preference for a similar interface in acupuncture needle insertion. To gain deeper insights into why MRUCT was favored in clinical settings (RQ4), we surveyed the 30 participants who chose MRUCT, asking them to highlight the specific advantages they found using an additional questionnaire.

Among the three doctors who participated, all appreciated the MRUCT’s adaptive UI positioning, its helpful insertion guidance, and its intuitive design. One doctor specifically valued the flexibility of manually adjusting reference insertion trajectories—a feature of our system that enhances user control—while two believed that the system facilitated their understanding of acupoints.

Of the 27 students, 25 were impressed with the MRUCT’s adaptive UI, which dynamically adjusts based on the location of the needle tip. Features such as insertion guidance, intuitive design, and the ability to learn acupoints also received high marks. However, fewer students considered the system's extensive features as beneficial to their experience.

\begin{figure}[htbp]
    \centering
    \includegraphics[width=\linewidth, height = 1.2in]{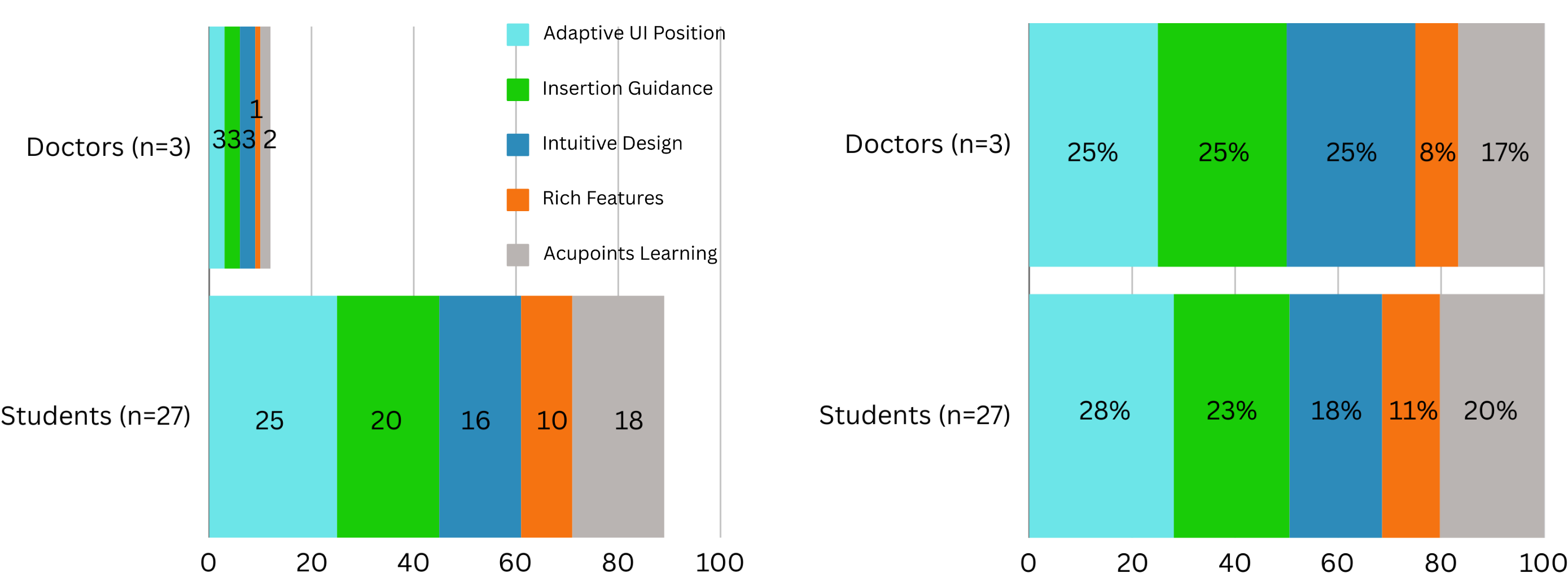}
    \caption{Specification of potential benefits of MRUCT in clinical settings, categorized by doctors and students. The left figure shows the number counts and the right figure illustrates values in percentile.}
    \label{fig:final}
    \vspace{-1.0em}
\end{figure}

Comparing the preferences of doctors with students (see \cref{fig:final}), a higher percentage of doctors enjoyed the intuitive design of MRUCT (7\% higher). A higher percentage of students believe that MRUCT can help with their learning of acupoints (3\% higher). The distribution of other potential benefits among each group remains similar. Since we only enrolled three doctors, incorporating more doctors into the study would improve our understanding of MRUCT’s benefits.

\subsubsection{Quantitative Results}
\ADD{Precision in acupuncture is critical for ensuring therapeutic efficacy and patient safety \cite{molsberger2012acupuncture}. Studies indicate that an insertion accuracy within 1-3 mm is generally sufficient for most acupuncture points, particularly when targeting muscle or connective tissue layers \cite{vickers2018acupuncture}. This standard ensures that needles are inserted with sufficient accuracy to avoid damage to surrounding structures while maximizing therapeutic effects.} 
Using the method outlined in the previous subsections for comparing acupuncture accuracy, we obtained the following quantitative analysis data:
\begin{table*}[htbp]
\centering
\caption{Quantitative analysis comparing three navigation methods—Finger-Cun, MRUCT, and Two-ring—among two experienced practitioners.}
\begin{tabular}{ccccc}
\hline
Users & Navigation Method & Entry Point Error(mm) & End Point Error(mm) & Total Time to Complete (sec.) \\ \hline
\multirow{3}{*}{G1} & Finger-Cun   & 0.72±0.07          & 1.25±0.10          & 214s         \\
                    & MRUCT (ours) & \textbf{0.49±0.06} & \textbf{0.79±0.11} & \textbf{67s} \\
                    & Two-ring     & 0.61±0.09          & 1.03±0.08          & 81s          \\ \hline
\multirow{3}{*}{H1} & Finger-Cun   & 0.80±0.02          & 1.44±0.07          & 258s         \\
                    & MRUCT (ours) & \textbf{0.52±0.05} & \textbf{0.81±0.09} & \textbf{72s} \\
                    & Two-ring     & 0.65±0.07          & 0.95±0.10          & 80s          \\ \hline
\end{tabular}
\label{tab:R2}
\end{table*}

As shown in \cref{tab:R2}, the MRUCT method performs best, minimizing errors in both the "entry point error" and "end point error" accuracy metrics. For both G1 and H1 users, the MRUCT method significantly reduces point localization errors and outperforms the other two methods in terms of operation time \ADD{while these results align closely with the clinical benchmarks}, making it the most efficient. This demonstrates its potential to help experienced practitioners perform acupuncture tasks more effectively \ADD{and utility for safe and effective acupuncture practice}.
\begin{figure}[htbp]
    \includegraphics[width=1.0\linewidth]{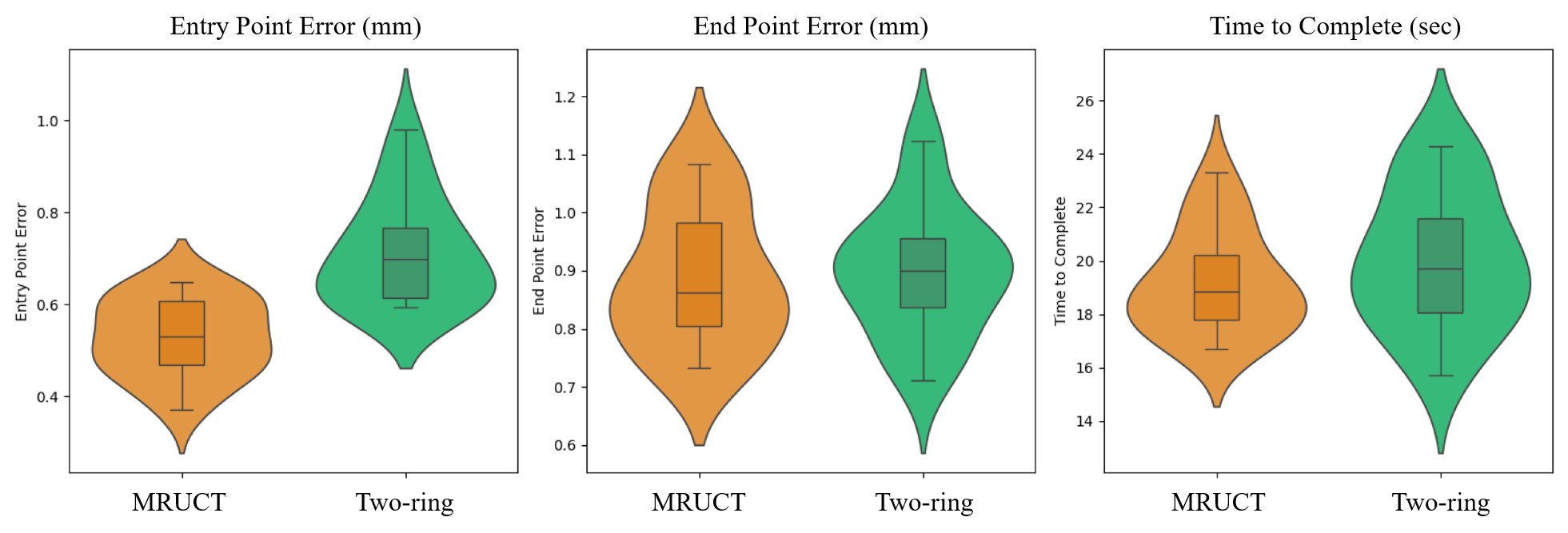}
    \caption{Quantitative analysis comparing two MR navigation methods—MRUCT and Two-ring—among six medical students with limited acupuncture experience.}
    \label{fig:RQ3}
    \vspace{-1.0em}
\end{figure}

As we can see in \cref{fig:RQ3}, the MRUCT method demonstrates a smaller error distribution for both the Entry Point Error and End Point Error metrics, showing a lower median and narrower error interval in the box-and-whisker plot. In contrast, the Two-ring method exhibits a wider error distribution and a higher median, indicating that MRUCT provides superior acupuncture guidance. Additionally, the MRUCT method shows a lower median and narrower distribution for the Time to Complete metric, suggesting that it requires less operation time and has higher user acceptance. Overall, MRUCT delivers greater accuracy and task efficiency for acupuncture guidance, making it a superior system.

\section{Discussion}
Overall, the findings of this study demonstrated the utility and versatility of MRUCT in enhancing the acupuncture guidance workflow. It successfully addressed the three research questions:

For RQ1, MRUCT utilizes the LDDMM algorithm to facilitate the registration of the template image with the target image while ensuring the consistency of anatomical structures associated with acupoints before and after alignment, as well as the smoothness of the deformation field. Through questionnaires and face-to-face interviews, acupuncturists indicated that the bones were correctly aligned, but some muscle boundaries appeared distorted.

For RQ2, MRUCT renders both the acupuncture needle and the patient's anatomical structures in MR, helping experienced practitioners efficiently position \ADD{acupoints}\DEL{acupuncture points according to the intrinsic anatomical structures}. \DEL{Additionally, it accelerates the acupuncture treatment process by allowing fine-tuning of the needle trajectory}\ADD{It also allows fine-tuning of needle trajectories, accelerating the acupuncture process}. Without MRUCT, \DEL{acupoint localization accuracy was slightly reduced, and the time required to complete the acupuncture task was significantly increased for acupuncturists}\ADD{localization accuracy slightly decreased, and the time to complete the task increased significantly for acupuncturists}.

For RQ3, MRUCT can automatically generate reference trajectories for medical students with minimal training, based on real trajectories scanned from acupuncturists\DEL{ needling on the template patient}. It also allows students to adjust these auto-generated trajectories. Additionally, MRUCT features a 3DUI that replaces the traditional needle indication UI (two-ring). Both the questionnaire results and the analysis of trajectory error data indicate that MRUCT can help medical students with limited practical experience perform acupuncture tasks more effectively.

When considered as a whole, MRUCT is demonstrably superior to other 3D navigation systems in terms of clinical applicability and ease of use.

\subsection{Limitation and Future Work}
One limitation of the user study is the small number of experienced acupuncturists in the test user group \ADD{and the lack of diverse patients in the clinical validation}, making direct comparisons with medical students challenging and hindering a full demonstration of the system's potential in aiding acupuncture navigation. Additionally, \ADD{while spatial alignment is effective for optimizing entry points, its reliability diminishes as the needle advances into deeper tissues. This decline in precision is influenced by factors such as soft tissue deformation, needle deflection, and variability in tissue elasticity. Furthermore, }by considering the entire muscle as a whole, we did not account for changes in the patient's preoperative and intraoperative muscle morphology. This prevented the acupuncturist from accurately locating acupuncture points based on the active muscle.

In the future, we plan to optimize the entire system in \DEL{three}\ADD{four} areas: (1) collecting more experimental data from experienced acupuncturists to refine the UI design according to the specific needs of doctors \ADD{while expanding the clinical validation across diverse patients to demonstrate the generalizability of MRUCT}; (2) researching the interaction between the needle tip in the virtual environment and the anatomical structure after alignment, as well as incorporating a force-feedback device to enhance the realism of needling; \DEL{and}(3) \ADD{incorporating lightweight pre-computation methods and} reducing system time delay while introducing a standard human body model as a pre-training tool for acupuncture, broadening the potential applications of MRUCT \ADD{and enhancing its practicality for more dynamic clinical workflows}; \ADD{and (4) integrating real-time muscle deformation models and adaptive imaging feedback into the MRUCT to ensure more accurate guidance}.
\section{Conclusion}
This paper presents MRUCT, a novel MR-based acupuncture guidance system that incorporates personalized ultrasonic computed tomography. In a formative study, we explored the needs of practitioners for a guided acupuncture procedure, identified challenges they face in daily practice, and derived insights into design choices for a well-integrated, user-friendly, and effective system.
Feedback from practitioners in the user study indicated that MRUCT is easy to use, informative, and intuitive, highlighting its potential to benefit medical students with limited practical experience in needling tasks.
By integrating medical imaging into the guidance system, MRUCT also uses LDDMM to complete UCT image registration, allowing practitioners to visualize intrinsic anatomical structures and adjust auto-generated trajectories.
Positive feedback and user ratings indicate a promising future for MRUCT as an invaluable tool for assisting with acupuncture tasks.
With further advancements and refinements, MRUCT holds the potential to revolutionize acupuncture guidance and contribute significantly to the development of acupuncture practitioners.


\acknowledgments{
The authors give special thanks to all the participants. This work has been supported by the National Key Technologies R\&D Program under Grants No. 2024YFB4708802, the NSFC under Grants No.62133009 and 92148205, the Natural Science Foundation of Jiangsu Province Major Project under grant BK20232008, Jiangsu Key Research and Development Plan under Grant  BE2023023-4 the Joint Fund Project 8091B042206 and the Fundamental Research Funds for the Central Universities.}

\bibliographystyle{abbrv-doi-hyperref-narrow}

\bibliography{template}
\clearpage
\listofchanges
\end{document}